\documentclass[amsmath,amssymb,reprint,aps,prl,floatfix,superscriptaddress]{revtex4-1}
\usepackage{bm}
\usepackage{amsmath}
\usepackage{graphicx}
\usepackage{braket}
\usepackage{bbm}
\usepackage[hidelinks]{hyperref}
\usepackage{makecell}
\begin{document}

\title{Single-shot discrimination of coherent states beyond the standard quantum limit}

\author{G.S. Thekkadath}
\email{g.thekkadath@imperial.ac.uk}
\affiliation{Department of Physics, Imperial College London, Prince Consort Rd, London SW7 2AZ, UK}
\affiliation{National Research Council of Canada, 100 Sussex Drive, Ottawa, Ontario K1A 0R6, Canada}

\author{S. Sempere-Llagostera}
\affiliation{Department of Physics, Imperial College London, Prince Consort Rd, London SW7 2AZ, UK}

\author{B.A. Bell}
\affiliation{Department of Physics, Imperial College London, Prince Consort Rd, London SW7 2AZ, UK}

\author{R.B. Patel}
\affiliation{Department of Physics, Imperial College London, Prince Consort Rd, London SW7 2AZ, UK}
\affiliation{Clarendon Laboratory, University of Oxford, Parks Road, Oxford, OX1 3PU, UK}

\author{M.S. Kim}
\affiliation{Department of Physics, Imperial College London, Prince Consort Rd, London SW7 2AZ, UK}

\author{I.A. Walmsley}
\affiliation{Department of Physics, Imperial College London, Prince Consort Rd, London SW7 2AZ, UK}

\begin{abstract}
The discrimination of coherent states is a key task in optical communication and quantum key distribution protocols. 
In this work, we use a photon-number-resolving detector, the transition-edge sensor, to discriminate binary-phase-shifted coherent states at a telecom wavelength.
Owing to its dynamic range and high efficiency, we achieve a bit error probability that unconditionally exceeds the standard quantum limit (SQL) by up to 7.7 dB.
The improvement to the SQL persists for signals containing up to approximately seven photons on average and is achieved in a single shot (i.e. without measurement feedback), thus making our approach compatible with larger bandwidths.
\end{abstract}

\maketitle

Optical coherent states are ideal information carriers since laser light can be modulated at high speeds and can propagate in fibers over long distances with little distortion. 
In addition to many other degrees of freedom of light (e.g. spatial, frequency, polarization), information can be encoded in the intensity and phase of the coherent state.
For example, the binary-phase-shifted coherent states $\ket{\pm \alpha}$ encode a bit.
Since these are non-orthogonal quantum states, there are fundamental bounds on the ability for a receiver to determine the value of the bit.
This ensures the security of certain quantum cryptography protocols~\cite{bennett1992quantum,scarani2009security} but also limits the communication rate of channels restricted to weak signals such as in deep space optical links~\cite{boroson2018achieving}.

A classical receiver typically employs homodyne detection.
This measures the Gaussian distribution of quadrature amplitudes, with a mean value that depends on the phase of the local oscillator with respect to the signal. 
For a fixed phase that maximizes the distance between the two signal values, the overlap between the two Gaussian measurement distributions results in a bit discrimination error probability of 
\begin{equation}
    P_{\mathrm{SQL}} = \left(1-\mathrm{erf}(\sqrt{2}|\alpha|) \right)/2
    \label{eqn:sql}
\end{equation}
where $\mathrm{erf}(x) = 2/\sqrt{\pi} \int_0^x \exp{(-t^2)} dt$. 
We will refer to Eq.~\eqref{eqn:sql} as the standard quantum limit (SQL).
If the signal intensity $|\alpha|^2$ is constrained, the error probability can be significantly reduced by measuring the signal in the photon-number basis.
In 1973, Kennedy proposed first displacing the signal then measuring the displaced signal $\hat{D}(\beta)\ket{\pm \alpha}$ using a click detector (i.e. non-photon-number resolving)~\cite{kennedy1973near}.
The local oscillator amplitude $\beta$ performing the displacement can be optimized to minimize the discrimination error and tends to $\beta=\alpha$ for large signal amplitudes~\cite{takeoka2008discrimination,wittmann2008demonstration,tsujino2011quantum}.
By updating $\beta$ during the measurement process using a feedback loop, the so-called Dolinar receiver can -- in principle -- reach the ultimate limit in discrimination error, 
\begin{equation}
    P_{\mathrm{Hel}} =\left(1-\sqrt{1-\exp(-4|\alpha|^2})\right)/2
    \label{eqn:helstrom}
\end{equation}
i.e. the Helstrom bound~\cite{helstrom1976quantum,dolinar1973optimum,cook2007optical,becerra2013experimental,izumi2020experimental}.

In practice, experimental imperfections such as mode mismatch between the local oscillator and the signal~\cite{li2013suppressing,becerra2015photon,dimario2018robust} or phase diffusion~\cite{bina2017homodyne,dimario2019optimized,dimario2020phase} lead to imperfect interference and hence increased discrimination error.
In such cases, photon-number-resolving detectors can significantly improve the discrimination capabilities of a Kennedy or Dolinar receiver by providing more information about the displaced signal than a binary click detector.
Previous experimental demonstrations achieved photon-number-resolution by employing a click detector and detecting signals whose time duration was much longer than the detector deadtime, thus reducing the receiver bandwidth~\cite{wittmann2010demonstration,becerra2015photon,dimario2018robust}.
Likewise, most demonstrations achieving sub-SQL error rates implemented the Dolinar receiver~\cite{cook2007optical,becerra2013experimental,becerra2015photon,izumi2020experimental} whose bandwidth is ultimately limited by the measurement feedback electronics. 
These approaches cannot be scaled to the large bandwidths required in applications like communications.
On the other hand, intrinsically number-resolving detectors such as transition-edge sensors (TESs)~\cite{lita2008counting} and superconducting nanowires~\cite{cahall2017multi,zhu2020resolving,endo2021quantum} count photons in a single shot and thus can operate at faster rates and without complex feedback electronics.
A single-shot receiver using a TES was demonstrated in Ref.~\cite{tsujino2011quantum}, but it did not use the photon-number information of the detector and achieved a small improvement to the SQL of 0.2 dB.

In this work, we employ a photon-number-resolving TES detector in a Kennedy receiver to discriminate binary-phase-shifted coherent states.
Owing to the high efficiency and single-photon sensitivity over a large dynamic range of the TES, we demonstrate an unconditional improvement to the SQL for signals containing up to approximately seven photons on average.
Our receiver achieves up to a 7.7 dB improvement to the SQL which is by far the largest achieved without measurement feedback.
Moreover, it operates at a telecom wavelength (1550 nm) that is compatible with fiber communication networks.

\begin{figure}
    \centering
    \includegraphics[width=1\columnwidth]{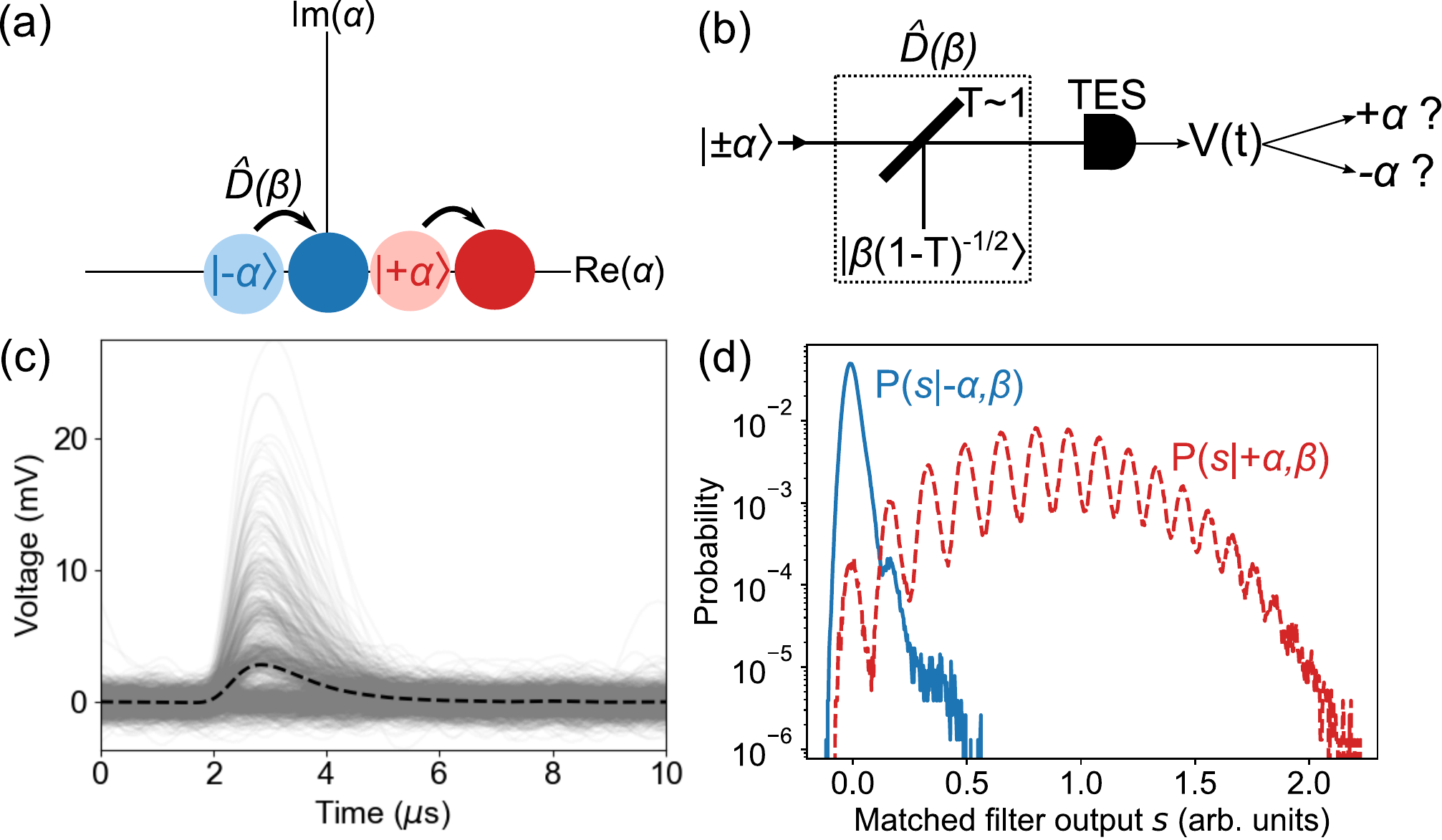}
	\caption{(a) Phase-space diagram. The receiver displaces $\ket{-\alpha}$ towards the vacuum state and $\ket{+\alpha}$ away from the vacuum state in order to make the two signals maximally distinguishable in the photon-number basis.
	(b) The displacement operation $\hat{D}(\beta)$ is approximated by combining the signal with a local oscillator $\ket{\beta/\sqrt{1-T}}$ on a highly transmissive beam splitter.
	The displaced signal, $\ket{\pm \alpha + \beta}$, is then measured using a photon-number-resolving transition-edge sensor (TES) whose detection trace $V(t)$ is used to determine whether the signal was $+\alpha$ or $-\alpha$. 
	(c) Plot shows 1000 typical detection traces. Each trace [grey lines] is converted into a scalar value using $s=\int dt V(t)V_0(t)$ where the matched filter $V_0(t)$ [dashed black line] is obtained by averaging over many detection traces prior to data acquisition.
	(d) Example of the measured probability distributions $P(s|\pm \alpha, \beta)$ for $|\alpha|^2 = 1.50$ and $|\beta|^2=1.51$. 
	The overlap between the distributions determines the discrimination error [Eq.~\eqref{eqn:disc_err}].
	}
	\label{fig:concept}

\end{figure}

The concept of the receiver is shown in Fig.~\ref{fig:concept}.
A displacement operation of amplitude $\beta$ is approximated by combining the signal with a coherent state $\ket{\beta/\sqrt{1-T}}$ on a beam splitter of high transmissivity $T$.
The field in the transmitted port, $\hat{D}(\beta)\ket{\pm \alpha}=\ket{\pm\alpha+\beta}$, is then detected by a TES.
The shape of the detection trace $V(t)$ contains information about the number of photons absorbed [Fig.~\ref{fig:concept}(c)].
It is not efficient to directly use $V(t)$ in a discrimination criterion since the dimension of the space of possible traces is enormous.
In general, one employs some strategy to extract information about the number of absorbed photons from $V(t)$, e.g. the trace height~\cite{tsujino2011quantum}, rise time~\cite{cahall2017multi}, or a principle component analysis~\cite{humphreys2015tomography}.
For TESs, a matched filter approach works well since the traces have approximately the same shape but are scaled depending on the number of absorbed photons.
We employ this strategy in this work.
The matched filter $V_0(t)$ is prepared prior to data acquisition by averaging over many detection traces obtained using a signal containing a few photons on average.
During data acquisition, we calculate the inner product of each detection trace with the matched filter to produce a scalar value that quantifies the amount of energy absorbed, i.e. $s=\int dt V(t)V_0(t)$.
While $s$ can be binned into a photon number [see oscillations in Fig.~\ref{fig:concept}(d)], determining these bins becomes problematic at larger ($\gtrsim 15$) photon numbers where the TES no longer has single-photon energy resolution.
Rather, we will show that $s$ can be directly used in the discrimination criterion.
We note that previous work~\cite{gerrits2012extending} demonstrated a linear dependence between $s$ and signal intensities up to $10^6$ photons and hence $s$ contains information about the absorbed energy even when the TES is beyond its number-resolving regime.

We employ the maximum a posteriori probability discrimination criterion.
That is, given the outcome $s$ and displacement amplitude $\beta$, the receiver assigns the value $+\alpha$ if
\begin{equation}
     P(+\alpha | s, \beta) > P(-\alpha | s, \beta),
\end{equation}
and otherwise it assigns the value $-\alpha$.
With this criterion, it can be shown that the discrimination error is given by~\cite{dimario2018robust} 
\begin{equation}
    P_{err} = 1 - \frac{1}{2}\sum_{s} \max_{\pm} \left[ P(s|\pm \alpha, \beta) \right]. 
\label{eqn:disc_err}
\end{equation}
We assumed equal prior probability for both $\pm\alpha$.
The probability $P(s|\pm \alpha, \beta)$ can be determined using the positive operator-valued measure $\{ \hat{\Pi}^s \}$ (POVM) of the TES:
\begin{equation}
    P(s|\pm \alpha, \beta) = \braket{(\pm \alpha+\beta) |\hat{\Pi}^s|(\pm \alpha+\beta)}.
\end{equation}
In general, the POVM element $\hat{\Pi}^s$ is a statistical mixture of photon-number states which is unknown to the experimentalist.
Although it can be reconstructed using detector tomography~\cite{humphreys2015tomography}, this requires numerous measurements using a calibrated coherent state source.
Alternatively, the optimal $\beta$ that minimizes Eq.~\eqref{eqn:disc_err} can be determined empirically which is the approach we employ in our experiment.

\begin{figure}
    \centering
    \includegraphics[width=1\columnwidth]{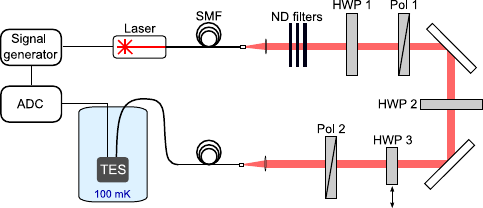}
	\caption{Experimental setup. Details provided in the main text. SMF: single-mode fiber, ND: neutral density,
	HWP: half-wave plate, Pol: Glan-Taylor polarizer, TES: transition-edge sensor, ADC: analogue-to-digital converter.
	}
	\label{fig:expSetup}

\end{figure}

The experimental setup is shown in Fig.~\ref{fig:expSetup}.
A signal generator drives a fiber-coupled laser diode (wavelength 1550 nm) which produces 1 $\mu$s square light pulses at a repetition rate of 60 kHz.
The pulses are attenuated using neutral-density filters. 
Two modes are encoded in polarization (H: horizontal, V: vertical) which provides common-path phase stability between the signal and local oscillator modes.
A half-wave plate (HWP 1) and polarizer (Pol 1) prepare $\ket{\gamma}_H\ket{0}_V$ where $|\gamma|^2 = |\alpha|^2 + |\beta|^2/(1-T)$.
Then, the angle of HWP 2 is adjusted to prepare the state $\ket{\alpha}_H\ket{\beta/\sqrt{1-T}}_V$.
To prepare $\ket{-\alpha}_H\ket{\beta/\sqrt{1-T}}_V$, we insert HWP 3 with its fast axis along $V$.
Pol 2 has transmissivity $T = 0.982(2)$ for $H$-polarized light.
The transmitted signal is measured using a TES held at a temperature of $\sim 100$ mK inside a dilution refrigerator.
The voltage trace produced by the TES is recorded using an analogue-to-digital converter triggered by the signal generator.
We perform the matched-filter calculation on batches of 1500 detection traces in order to continuously acquire data while the calculation is performed.
The bandwidth of our receiver (60 kHz) is limited by the thermal recovery time of the TES.
Generally TESs are limited to bandwidths of hundreds of kilohertz since they are designed for high energy resolution and so use a weak thermal link to a heat sink to dissipate the absorbed energy~\cite{lita2008counting}. 

Before taking data, we prepare a look-up table which provides the wave plate angles needed to produce the desired values of $|\alpha|^2$ and $|\beta|^2$.
This table is created by treating the TES as a photon counter, i.e. binning the $s$ values into photon numbers and recording the average photon number measured for various wave plate angles.
The wave plates are mounted on motorized stages allowing us to set their fast axis to an accuracy of $0.1$ degrees.
This accuracy leads to an uncertainty on the order of 1\% on the preparation of $|\alpha|^2$ and $|\beta|^2$.
By measuring the average photon number as a function of the HWP 2 angle, we determine the interference visibility to be $\xi = 0.998(1)$ which is mainly limited by polarization-ellipticity introduced by the wave plates.
We determined in a previous work that our TES has an efficiency of $\eta=0.98^{+0.02}_{-0.08}$~\cite{humphreys2015tomography}.

\begin{figure}
    \centering
    \includegraphics[width=0.70\columnwidth]{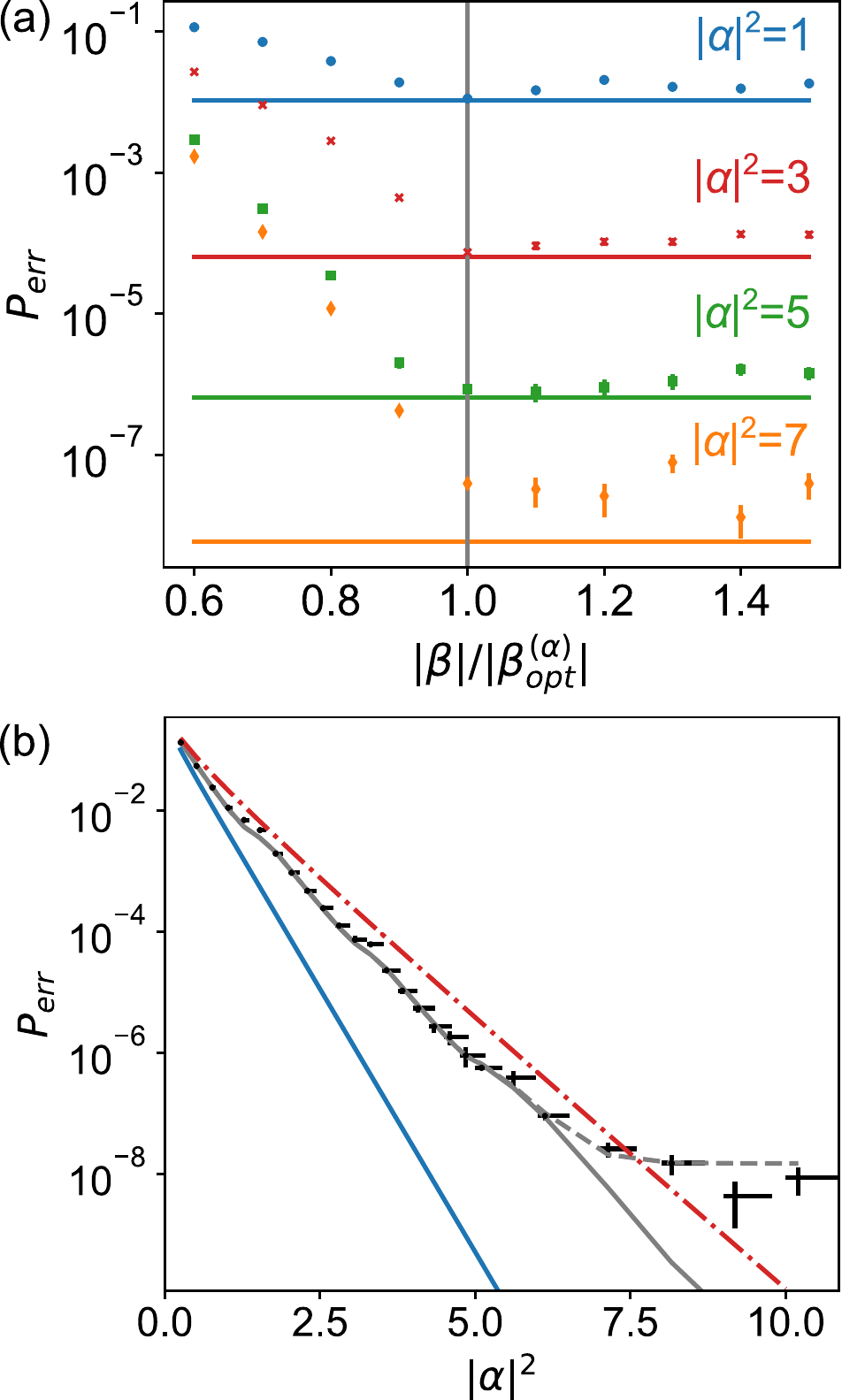}
	\caption{(a) Error probability $P_{err}$ as a function of the displacement amplitude $|\beta|$ for various signal intensities $|\alpha|^2$. 
	Solid horizontal lines are the expected error probabilities assuming ideal photon counting with $\xi = 0.9985$ and the optimal displacement $\beta^{(\alpha)}_{opt}$.
	(b) $P_{err}$ as a function of $|\alpha|^2$.
	Our measured error [black markers] agrees well with the expected error for ideal photon counting [grey line] up to $|\alpha|^2 \sim 6$.
	Beyond this signal intensity, dark counts cause $P_{err}$ to plateau [dashed grey line].
	Also shown is the standard quantum limit achievable with perfect detection efficiency  [Eq.~\eqref{eqn:sql}, red dot-dashed line] and the Helstrom bound [Eq.~\eqref{eqn:helstrom}, blue solid line].
	All error bars show one standard deviation. 
	Y-axis error bars are from Poissonian counting statistics.
	X-axis error bars are from the uncertainty in the receiver efficiency $\eta$.
	}
	\label{fig:results}

\end{figure}

We begin by determining the optimal displacement for each signal intensity.
We measure $P(s|\pm\alpha,\beta)$ for each $\beta$ using at least $10^6$ trials and determine the discrimination error $P_{err}$ using Eq.~\eqref{eqn:disc_err}.
In Fig.~\ref{fig:results}(a), we plot $P_{err}$ as a function of $|\beta|/|\beta^{(\alpha)}_{opt}|$ for various signal intensities.
Here, $\beta^{(\alpha)}_{opt}$ is the expected optimal displacement for $\alpha$ using ideal photon counting and is calculated in the following manner.
Given $\alpha$, we compute $P_{err}$ using
\begin{equation}
\begin{split}
P_{err} &= 1 - \frac{1}{2}\sum_{n} \max_{\pm} \left[ P(n|\pm \alpha, \beta) \right]\\
&= 1 - \frac{1}{2}\sum_{n} \max_{\pm} \left[ (N_\pm)^n e^{-N_\pm} / n! \right]
\end{split}
\label{eqn:expected_error}
\end{equation}
where $N_\pm = T|\alpha|^2 + |\beta|^2 \pm 2\xi \sqrt{T}|\alpha||\beta|$.
This uses the same discrimination criterion as in Eq.~\eqref{eqn:disc_err} but we have replaced $P(s|\pm \alpha, \beta)$ with the displaced signal photon-number distribution. 
Then, $\beta^{(\alpha)}_{opt}$ is determined by numerically minimizing $P_{err}$.
As can be seen from the data points along the grey vertical line in Fig.~\ref{fig:results}(a), the optimal displacement for our TES corresponds to $\beta^{(\alpha)}_{opt}$ despite not binning the detector output into a photon number.
The error probability obtained using this optimal displacement is plotted as a function of the signal intensity in Fig.~\ref{fig:results}(b) and agrees with the expected error for an ideal photon counter with $\xi = 0.9985$ [grey line] up to $|\alpha|^2 \sim 6$.
This suggests that our matched filter technique is extracting all the relevant information about the displaced signal (i.e. its energy) from the detection traces.
For $|\alpha|^2 > 7$, the measured $P_{err}$ plateaus.
This is not due to the finite number resolution of the TES: one only requires the ability to count up to 6 photons for $|\alpha|^2 < 10$, as discussed in Ref.~\cite{dimario2018robust}, whereas TESs can faithfully count beyond 10 photons.
Rather, we attribute this plateau to detector dark counts.
Although the dark count probability is about $10^{-3}$ per pulse, most of these dark counts produce small $s$ values which do not overlap with $P(s | +\alpha, \beta)$ at large $|\alpha|^2$ and hence do not limit $P_{err}$.
High-energy dark counts (i.e. greater than 15 photons) which do overlap with $P(s | +\alpha, \beta)$ at high $|\alpha|^2$ occur with a probability of about $P_{\mathrm{dark}} \sim 10^{-8}$ per pulse and are likely caused by cosmic rays~\cite{dreyling2015characterization}.
The dashed grey line in Fig.~\ref{fig:results}(b) shows the expected error in the presence of such dark counts.
It is obtained by adding $P_{\mathrm{dark}} = 3\times 10^{-8}$ to $ P(n|\pm \alpha, \beta)$ for $n>15$, re-normalizing the distribution, and calculating $P_{err}$ using Eq.~\eqref{eqn:expected_error}. 

To compare the performance of our receiver to an idealized homodyne receiver with perfect efficiency [red dot-dashed line], the measured $|\alpha|^2$ is rescaled to $|\alpha|^2/\eta$ where $\eta=0.98^{+0.02}_{-0.08}$ is the probability that a photon inside the optical fiber attached to the TES is detected~\cite{humphreys2015tomography}.
We find that our receiver outperforms this idealized homodyne detector up to $|\alpha|^2 \sim 7.5$.
The largest improvement is $7.7^{+0.9}_{-3.9}$ dB obtained at $|\alpha|^2 = 4.8$. %
In Table~\ref{table:performence}, we compare the performance of our receiver to others that also achieve an unconditional improvement to the SQL.
As with those works, we consider free-space-to-fiber coupling losses to be part of the state preparation and hence these are not included in the receiver efficiency $\eta$.

\begin{table}
\begin{center}
\resizebox{\columnwidth}{!}{%
\begin{tabular}{ c c c c c c}
\hline
 Reference & \makecell[c]{Improvement\\to SQL (dB)} & Single shot? &  \makecell[c]{Repetition\\rate (Hz)} &  \makecell[c]{Wavelength \\ (nm)} \\ 
 \hline
  \cite{tsujino2011quantum} & 0.2 & Yes & $4 \times 10^4$ & 853 \\
  \cite{cook2007optical} & 0.4 & No & $5 \times 10^4$ & 852 \\
 \cite{izumi2020experimental} & 1.2 & No & $5 \times 10^3$ & 1550 \\
  \cite{becerra2013experimental} & 6.5 & No & $1.25 \times 10^4$ & 633 \\
 This work & 7.7 & Yes & $6 \times 10^4$  & 1550 \\    
  \cite{becerra2015photon} & 14.5 & No & $1.1 \times 10^4$ & 633 \\
  \hline
\end{tabular}%
}
\end{center}
\caption{Receivers achieving an unconditional improvement to the SQL.}
\label{table:performence}
\end{table}

In summary, we built a coherent-state receiver using a transition-edge sensor detector to discriminate a binary-phase-shifted signal in a single shot.
We observe a bit error probability below the standard quantum limit with signals containing up to seven photons on average owing to the high efficiency and dynamic range of the transition-edge sensors.
Our results also demonstrate that the continuous output of a photon-number-resolving detector (such as the trace area) can be directly used in a discrimination criterion without having to convert it into a photon number.
This approach can simplify and potentially speed-up the trace analysis required in a receiver employing fast number-resolving detectors such as superconducting nanowires which have bandwidths in the tens of megahertz range~\cite{cahall2017multi,zhu2020resolving,endo2021quantum}. 
Minimizing the analysis time is especially important when the measurement outcome is used in a feedback loop like in the Dolinar receiver.

\begin{acknowledgements}
We thank K. Banaszek for the discussions that prompted this work.
This work was supported by: Engineering and Physical Sciences Research Council (P510257, T001062); H2020 Marie Skłodowska-Curie Actions (846073). 
\end{acknowledgements}




\bibliographystyle{apsrev4-1}
\bibliography{refs}

\end{document}